\documentstyle[jkas]{article}

\beginpage{1}
\endpage{13}
\year{2014}\volume{47}\month{February}
\runningauthor{H. B. ANN}
\runningtitle{{ENVIRONMENT DEPENDENCE OF DISK}}

\month{February} \year{2014} \volume{47} \issueno{1}
\begin{document}
\title{ENVIRONMENT DEPENDENCE OF DISK MORPHOLOGY OF SPIRAL GALAXIES}
\author{Hong Bae Ann}
\address{Department of Earth Science Education, Pusan National University, Pusan 609-735, Korea\\ {\it E-mail : hbann@pusan.ac.kr}}
\address{\normalsize{\it (Received October 27, 2013; Revised November 26, 2013; Accepted December 4, 2013)}}
%\offprints{H. B. Ann}
%--------------------------------------------------------------------
\abstract{
We analyze the dependence of disk morphology (arm class, Hubble type, bar type)
of nearby spiral galaxies on the galaxy environment by using local
background density ($\Sigma_{n}$), projected distance ($r_{p}$), and tidal
index ($TI$) as measures of the environment. There is a strong dependence
of arm class and Hubble type on the galaxy environment, while the bar type
exhibits a weak dependence
with a high frequency of SB galaxies in high density regions.
Grand design fractions and early-type fractions
increase with increasing $\Sigma_{n}$, $1/r_{p}$, and $TI$,
while fractions of flocculent spirals and late-type spirals decrease.
Multiple-arm and intermediate-type spirals exhibit nearly constant fractions
with weak trends similar to grand design and early-type spirals.
While bar types show only a marginal dependence on $\Sigma_{n}$, they show
a fairly clear dependence on $r_{p}$ with a high frequency of SB galaxies at
small $r_{p}$. The arm class also exhibits a stronger
correlation with $r_{p}$ than $\Sigma_{n}$
and $TI$, whereas the Hubble type exhibits similar correlations with
$\Sigma_{n}$ and $r_{p}$. This suggests that the arm class is mostly
affected by the nearest neighbor while the Hubble type is affected by the
local densities contributed by neighboring galaxies as well as the
nearest neighbor.
}
\maketitle

%--------------------------------------------------------------------

\section{INTRODUCTION}

The structure of a spiral galaxy is characterized by two primary components,
disk and bulge of which radial luminosity distributions are well represented
by an exponential profile \citep{fre70} and $r^{1/4}$ law
profile \citep{dev53}, respectively. The ratio of the luminosities in these
two components, the disk-to-bulge ratio, is a well defined quantity which
can be used as a proxy of Hubble type.
The spiral arms and bars are the secondary
components displaying perturbations imposed on the disk. However, they are as
important as the primary components, disk and bulge, because their structures
are closely related to the internal and external properties of the galaxies \citep{elm11} and they
are thought to drive secular evolution of the spiral galaxies \citep{kor79,pfe90,fri93,fri95,zha96,zha98,zha99, kor04}.

In Hubble's morphological classification of galaxies, spiral galaxies
are divided into subtypes, known as Hubble stages (early-, intermediate-
and late-type), according to the dominance of bulge luminosity, arm openness,
and patchness of the spiral arms. There are quantitative measures of the
first and second properties, namely, bulge-to-disk ratio and the pitch angle
of the spiral arms. They are considered to reflect the physical properties
of the disk morphology. But it is difficult to directly relate the patchness
with the physical properties of galaxies because it is affected  by the
distance of a galaxy.
Thus, recent morphological studies of spiral galaxies consider the
bulge-to-disk ratio and arm openness as the working criteria to distinguish
the Hubble stage of spiral galaxies. Early-type spirals have tightly wound
spiral arms and a large bulge-to-disk ratio, while the late-type spirals have
loosely wound spiral arms and a small bulge-to-disk ratio.

However, the openness of the spiral arms is not the only property of spiral
arm structures. \citet{elm82,elm87} introduced arm classes, which are
sometimes denoted as AC, to distinguish the arm structures on the basis of
the degree of symmetry and continuity together with the number of arms.
The 12 arm classes introduced by \citet{elm82} were reduced to 10 arm classes,
as two classes that are related to the presence of a bar (AC10) and
companion (AC11) were removed. However, most previous studies divided
the arm classes into two or three broad classes: (1) grand design (G) and
flocculent (F) or (2) grand design (G), multiple-arm (M), and flocculent (F).
For the case of two broad classes, multiple-arm is considered as a subclass
of the grand design arms. Grand design arms have two long symmetric arms,
while flocculent arms have a large number of chaotic fragmented arms.
Multiple arms exhibit intermediate properties between the grand design
and flocculent arms. Some multiple-arm galaxies have short
grand design arms in the inner parts and
multiple arms in the outer parts. Some of the outer arms in multiple-arm
galaxies are likely to be connected to produce pseudo-rings \citep{elm82}.

%%%%%%%%%%%  Fig.1 %%%%%%%%%%%%%%%%%%%%%%%%%%%%%%%%
\begin{figure*}[!t]
\center
\includegraphics[scale=0.90]{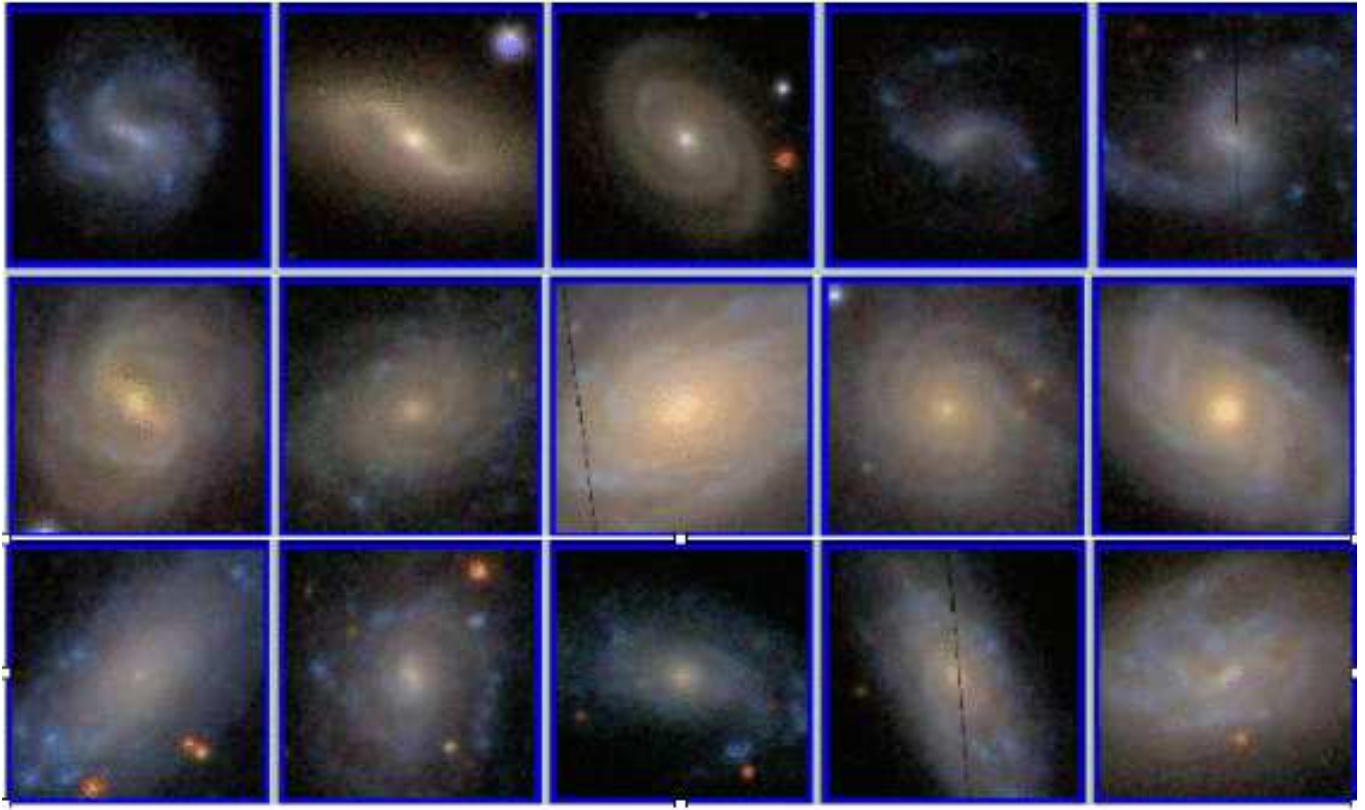}
\caption{Sample of spiral galaxies. Grand design galaxies, multiple-arm
galaxies and flocculent galaxies are displayed in the top panels, middle
panels and bottom panels, respectively.
}
\end{figure*}

There have been many investigations regarding the relationships between
arm classes and internal properties, including luminosity, color,
disk-to-bulge ratio, and morpholgy. The variation in the mean luminosity
and colors of galaxies across different arm classes is small~\citep{ann13}.
Conversely, considerable differences
are observed in the galaxy size~\citep{elm87,ann13} and  shape of
rotation curves~\citep{elm90,biv91}. There seems to be little or no dependence
of the arm classes on the star formation rate~\citep{elm86} and star formation
related quantities such as the neutral hydrogen content \citep{rom85},
CO surface brightness \citep{sta87}, and supernova rate \citep{mcc86}.

Arm classes exhibit some dependencies on the Hubble type \citep{elm82,cho11,
ann13}. Flocculent galaxies are more likely to occur in late-type galaxies,
while multiple-arm galaxies are more frequent in early-type galaxies. Grand
design galaxies are also more frequent in early-type galaxies, but the
correlation is much weaker than multiple-arm galaxies. There is a
correlation between the arm class and bar type (SA, SAB, and SB).
The fraction of
grand design galaxies increases from SA to SAB to SB, while
flocculent galaxies show the opposite correlation.
Multiple-arm galaxies exhibit nearly constant fractions regardless of
the bar type. If we consider the Hubble type and
bar type together, early-type SB galaxies are likely to have a high fraction of
grand design galaxies; however, late-type SB galaxies exhibit no such
tendency \citep{elm89}.

The environment dependence of galaxy morphology is well understood for field
galaxies \citep{got03,par07} as well as cluster galaxies \citep{dre80}. It
is represented by the morphology-density relation (MDR) or morphology-radius
relation (MRR). In other words, early-type galaxies (E/S0) are likely to be
found in the dense environment or central regions of clusters, while late-type
galaxies (Sp/Irr) are more frequently observed in the under-dense regions
or outer parts of clusters.
The environment dependencies of the disk morphology of spiral galaxies, which
are represented by
arm class, Hubble type, and bar type, is of interest
because the arm class is closely related to the driving mechanism of the spiral
structures \citep{elm86,elm11}. Grand design arms are thought to be formed
by density waves \citep{lin64,ber89a,ber89b}, whereas flocculent
arms are thought to be formed by stochastic self-propagated
star formation \citep{ger78, sei82}. Thus, understanding the
environment of galaxies with different arm classes provides some information
about the conditions for various modes of star formation.

\begin{table}
 \centering
%\begin{minipage}{140mm}
  \caption{Frequency of arm class, Hubble type and bar type of nearby spiral galaxies.}
  \setlength{\tabcolsep}{1.35mm}
  \begin{tabular}{@{}cccccccccc@{}}
  \hline
   & \multicolumn{3}{c}{arm class}  & \multicolumn{3}{c}{Hubble stage}  &
   \multicolumn{3}{c}{bar type} \\
  \hline
     &  G &   M &    F &    e &   i &   l &   SA & SAB & SB \\
 \hline
 N   &  347 &  656 &  909 &  382 &  893 &  637 &  639 &  536 &  630 \\
\hline
\end{tabular}
%\end{minipage}
\end{table}

Early investigations into the environment dependence of arm classes
\citep{elm82,eed82} found that grand design galaxies are
more frequent in high density regions, while flocculent galaxies
favor low density regions.
However, \citet{giu89} reported that flocculent galaxies are more
frequent in interacting or binary systems, which are likely to be
present in dense environments. These contradictory claims may be due to
poor statistics \citep{giu94}. A recent study of the arm class and local
background density of nearby spiral galaxies \citep{ann13} confirmed the
earlier findings of \citet{elm82}, \citet{eed82}
and \citet{elm87}. \citet{ann13} found that there is a weak trend of increase
in the fraction of grand design galaxies with increasing local background
density, while a stronger opposite trend of a decreasing fraction
with increasing density was found for flocculent galaxies.
They derived the local background density
from the projected distances to the 5th nearest neighbor.
However, as described by \citet{mul12}, the local background densities derived
using the $n$th nearest neighbor method depend on two parameters $\Delta V^{*}$
and $M^{*}$ which set the upper limit of the line of sight velocity difference
between the target galaxy and its neighbor and the lower limit of the
luminosity of the neighbor galaxies to be searched.
Thus, it is necessary to choose a set of
parameters ($n$, $\Delta V^{*}$, $M^{*}$) suitable for the local background
density and relevant to the formation of spiral structures.

The natural choice for $M^{*}$ appears to be the limiting magnitude that
defines the sample galaxies as a volume-limited sample. However, other values
of $M^{*}$ are also possible if the selected $M^{*}$ is relevant to a
physical process being investigated. In case of $\Delta V^{*}$, it would be
different for galaxies in different environments because peculiar
velocities depend on the large scale distribution of galaxies
as well as nearby galaxies. Moreover, \citet{par09} argued
that $\Delta V^{*}$ depends on the morphology of galaxies.
They adopted $\Delta V^{*}$=~400 and 600~km/s for late- and early-type
galaxies, respectively, to search for the nearest neighbor galaxies on the
basis of the pairwise velocity difference between the target galaxy
and their neighbors \citep{par08,par09}. Since we are dealing with
galaxies in a variety of environments, a proper value  of $\Delta V^{*}$ is
yet to be determined. In addition to this, it is quite likely that disk
morphology correlates well with other environment measures than local
background density.

The purpose of the present study is to determine which measures of
environment are closely related to the disk morphology.
To do this, we first explore the dependence of local background densities on
the neighbor search parameters, $M^{*}$, $\Delta V^{*}$, and $n$,
using the Sloan Digital Sky Survey (SDSS) Data Release 7 (DR7). Subsequently,
we study local background densities of spiral galaxies, for which
arm class, Hubble type, and bar type are known. We look for the
appropriate set of
$n$, $M^{*}$, and $\Delta V^{*}$ that we could use to derive the local
background densities relevant to the disk morphology.

In Section 2, we describe the observational data and method of arm
classification. Local background density dependence of disk morphology with
effects of neighbor search parameters on the local background densities
are presented in Section 3. We descrive other measures of local environment
in Section 4. Discussion on the relation between the disk morphology and the
local environment is given in Section 5 and conclusions of the present study
are given in Section 6 followed by acknowledgements in the last section.

%%%%%%%%%%%%%%%%%%%%%%%%%%%%%%%%%%%%%%%%%%%%%%%%%%%%%%%%%%%%%%%%%%%%%%%%%%

%%%%%%%%%%%%%%%%%%%%%%%%%%%%%%%%%%%%%%%%%%%%%%%%%%%%%%%%%%%%%%%%%%%%%%%%%%

\section{OBSERVATIONAL DATA}

The present sample of local galaxies was extracted from the Korea
Institute for Advanced Study Value Added Galaxy Catalog (KIAS VAGC), which
provides the basic data of galaxies such as coordinates, redshifts, magnitudes,
and colors along with the broad morphological types determined
by automated classifiers \citep{par05}. See \citet{cho10} for a detailed
description of the KIAS VAGC. Since the KIAS VAGC is based on
the SDSS DR7, the present sample of galaxies is a flux-limited sample with
a limiting magnitude of $r=17.7$. Thus, the galaxies with $z < 0.02$ in the
present sample (13762 galaxies) is complete up to $M_{r}\approx-16.1$.
The sample of spiral galaxies used for the analysis of the environment
dependence of disk morphology is a subsample of the present sample, for which
arm classes, Hubble types, and bar types are given in \citep{ann13}. The
numbers of arm class, Hubble type and bar type are listed in Table 1.
Bar types of 107 spiral galaixes were not determined due to high inclinations.
We present the sample images of arm classes in Fig. 1.
We used the parent sample of 13762 galaxies as the basic data set
for the environment analysis.
%%%%%%%%%%%%%%%%%%%%%%% Fig2 %%%%%%%%%%%%%%%%%%%%%%%%%%%%%%%%%%%%%%%%%%%%%%
\begin{figure}[!t]
\center
\includegraphics[scale=0.6]{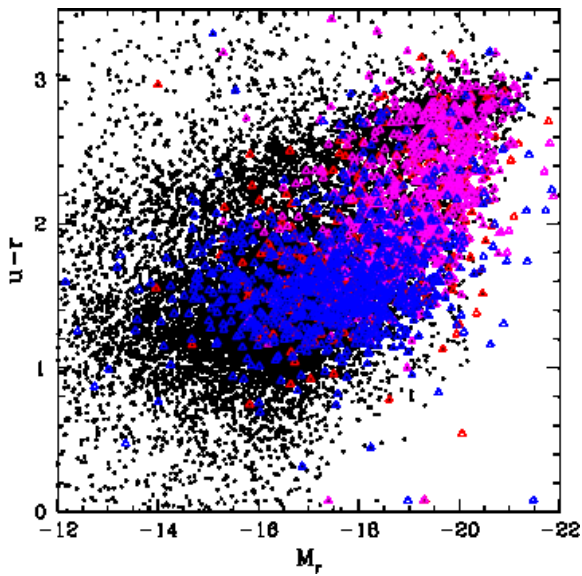}
\caption{$M_{r}$ versus $u-r$ plot showing the distribution
of nearby galaxies ($z < 0.02$) in KIAS VAGC. We designate spiral galaxies
with arm classes by colored triangles: G (red), M (magenta) and F (blue).
}
\end{figure}
%%%%%%%%%%%%%%%%%%%%%%%%%%%%%%%%%%%%%%%%%%%%%%%%%%%%%%%%%%%%%%%%%%%%%%%%%%
%%%%%%%%%%%%%%%%%%%%%%%%%%%%%%%%%%%%%%%%%%%%%%%%%%%%%%%%%%%%%%%%%%%%%%%%%%

%%%%%%%%%%%%%%%%%%%%%%%%%%%%%%%%%%%%%%%%%%%%%%%%%%%%%%%%%%%%%%%%%%%%%%%%%%

Fig. 2 shows the $u-r$ versus $M_{r}$ diagram of 13762 nearby galaxies.
In the figure, 1912 spiral galaxies are denoted as open triangles with colors
coded by their arm classes; G (grand design, red), M (multiple-arm, magenta)
and F (flocculent, blue).
The arm classes were based on the arm classification system of \citet{elm87}.
However, \citet{ann13} used three broad arm classes, grand design (G),
multiple-arm (M), and
flocculent (F), to suppress the statistical noise in the correlation studies
between the arm classes and other galaxy properties. Following the
arm class definition of \citet{elm87}, \citet{ann13} assigned AC 1-4 to
flocculent arms, AC 5-9 to multiple arms, and AC 12 to grand design arms.

\section{LOCAL BACKGROUND DENSITY AS A MEASURE OF ENVIRONMENT}

The local background density is one of the most widely used parameters
to represent the environment of a galaxy since \citet{dre80} determined the MDR
for cluster galaxies. There are several ways to define the local background
density. If we know the distances of the galaxies
accurately enough to determine the three-dimensional spatial
distribution of galaxies, we can calculate the spatial density
$\rho(r)$. However, except for the galaxies for which distances are
determined by the
Cepheid $P-L$ relation or Tully-Fisher relation \citep{tul77}, the distances
of galaxies are determined from the galaxy redshifts. Because the
redshifts of galaxies are affected by the peculiar motions of galaxies
due to the
interactions with neighbor galaxies or large scale structures,
the distances derived from the galaxy redshifts are not accurate enough,
especially
for galaxies in the local volume. To alleviate the effects of the inaccurate
distances on the derivation of the local background density, it is common
practice to derive a surface density $\Sigma$ centered on a target galaxy
using the projected distance $r_{p}$ to neighbor galaxies, which is
calculated from the distance of the target
galaxy and angular separation between the target galaxy and its neighbors.

There are two practical methods to calculate the surface density $\Sigma$
using $r_{p}$: one involves the use of fixed apertures and the other
involves the use of projected distances to the $n$th nearest neighbor galaxy.
The fixed aperture method with an aperture size of $5\sim8$~$~h^{-1}$Mpc
is known to be effective for densities at the superhalo scales,
while the $n$th nearest neighbor method with small $n$ is best suited for
densities inside high mass halos \citep{mul12}.  Hence, we adopt the
$n$th nearest neighbor method to derive the local background density in the
present study. The number of neighbors to be searched, $n$, can be selected by
considering the length scales relevant to the systems under investigation.
A small $n$ such as $n=3$ or $n=5$ is suitable for the local environment of
target galaxies in isolation or those in galactic groups, while a
large $n$ such as $n\geq9$ is suitable for the scale of clusters of galaxies.
However, even if $n$ or an aperture is selected, the local background density
of a galaxy cannot be determined uniquely because there are two
free parameters left to be determined: the limiting absolute magnitude of the
neighbor galaxies to be searched, $M^{*}$, and the maximum
velocity difference of the neighbor galaxies, $\Delta V^{*}$.

\subsection{Effect of ${M_{r}}^{*}$ and $\Delta V^{*}$ on Neighbor Search}

One of the most direct measurements of the galaxy environment is the projected
distance to the nearest neighbor ($r_{p}$). The morphology conformity between
nearest neighbors in galactic satellite systems \citep{ann08} and in
field galaxies \citep{par08} strongly depends on $r_{p}$.

All galaxy environment measures are required to define neighbors of a
target galaxy by constraining $M^{*}$ and $\Delta V^{*}$.
Since the projected distance ($r_{p}$) to the nearest neighbor is the most
direct measure of the galaxy environment, we examined the dependence of the
mean $r_{p}$
on $M^{*}$ and $\Delta V^{*}$ using nearby galaxies ($z < 0.02$)
listed in the KISG VAGC. We varied $\Delta V^{*}$ from 300 to 1000~km/s
with several $M^{*}$. The results of some typical cases are shown Fig.~3.
As shown in Fig.~3, the mean $r_{p}$ to the nearest
galaxy increases from the neighbor galaxies selected using a faint $M^{*}$ to
those selected by bright $M^{*}$. This is because the luminosity function
of local galaxies increases from bright galaxies to faint ones in the
luminosity ranges explored in Fig. 3.
Because $M^{*}=-16.1$ corresponds to the SDSS
observation limit of spectroscopic target galaxies at $z=0.02$,
the mean projected separation of galaxies in the
present volume limited sample is $\sim0.35$~$~h^{-1}$Mpc.

%%%%%%%%%%%%%%%%%%%%%%%%   fig3; fig3rpMr.eps  %%%%%%%%%%%%%%%%%%%
\begin{figure}[!t]
\centering
\includegraphics[scale=0.53]{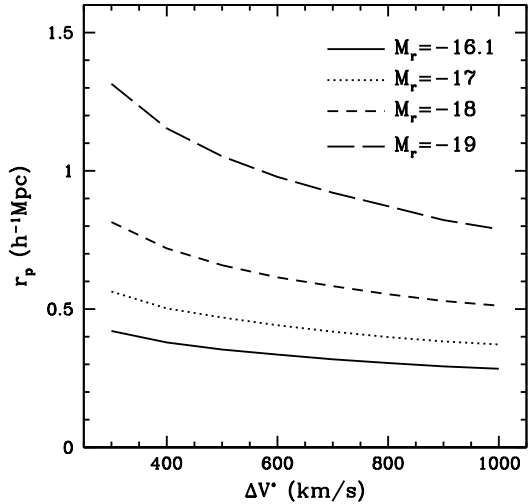}
%\vspace{-1.2cm}
\caption{Projected distance to the nearest neighbor $r_{p}$ as a function of
$\Delta V^{*}$. The line types distinguish the
luminosity constraints: $M^{*}=-16.1$ (solid line), $M^{*}=-17$ (dotted line),
$M^{*}=-18$ (short dashed line), and $M^{*}=-19$ (long dashed line).
}
\end{figure}
%%%%%%%%%%%%%%%%%%%%%%%%%%%%%%%%%%%%%%%%%%%%%%%%%%%%%%%%%%%%%%%%%%%%

For a given $M^{*}$, the mean $r_{p}$ decreases with increasing $\Delta V^{*}$,
regardless of $M^{*}$. However, the slopes of the variations are much different.
For the 300 to 1000~km/s $\Delta V^{*}$ range, the mean $r_{p}$
changes from $\sim0.3$ to $\sim0.4$ $~h^{-1}$Mpc for $M^{*}=-16.1$,
while it changes from $\sim0.8$ to  $\sim1.3$ $~h^{-1}$Mpc for
$M^{*}= -19$. The small changes in $r_{p}$ for the galaxies selected
by $M^{*}=-16.1$ is due to
presence of neighbor galaxies with $\Delta V$ smaller than 300~km/s within
$r_{p}\approx0.4$ $~h^{-1}$Mpc. The large $r_{p}$ for galaxies brighter
than ${M_{r}^{*}}=-18$ with $\Delta V$ less than 300~km/s is due to the absence
of bright companions as the closest neighbor. The mean $r_{p}$ of
$\sim1$~$~h^{-1}$Mpc is about three times larger than the virial radius of
a bright spiral galaxy such as the Milky Way.

Fig. 4 shows the mean $r_{p}$ of the $n$th nearest neighbor as a function of
$\Delta V^{*}$ for $M^{*}=-16.1$ and $M^{*}=-18$. Here, we show three cases
of $n=3$, $n=7$, and $n=21$. As shown in Fig. 4, the mean $r_{p}$
of the 3rd nearest neighbor with $M^{*}=-16.1$ increases very slowly from
$\sim0.5$ $~h^{-1}$Mpc for $\Delta V^{*}$=1000~km/s to $\sim$~1~$~h^{-1}$Mpc for
$\Delta V^{*}$ = 300~km/s.
The mean $r_{p}$ of the 3rd nearest neighbor with $M^{*}=-18$
is very similar to the mean $r_{p}$ of the 7th nearest neighbor with
$M^{*}=-16.1$. They vary slowly from $\sim1$ $~h^{-1}$Mpc to
$\sim1.3$ $~h^{-1}$Mpc for $\Delta V^{*}$ between 1000 and 500~km/s
and becomes $\sim$1.8 $~h^{-1}$Mpc for $\Delta V^{*}$=300~km/s.
A similar trend of rapid
increase of $r_{p}$ for $\Delta V^{*}$  $\lesssim500$ km/s is observed
for other values of $n$. Thus, there appears to be a common property, probably
caused by the paucity of companions for a small $\Delta V$. This suggests
that a $\Delta V^{*}$ of $\sim$~400~km/s is a lower limit
for searching neighbor
galaxies, regardless of the length scales to be studied.

%%%%%%%%%%%%%%%%%%%%%%%%   fig4; fig4rp1618.eps  %%%%%%%%%%%%%%%%%%%
\begin{figure}[!t]
\vspace{-0.4cm}
\center
\includegraphics[scale=0.53]{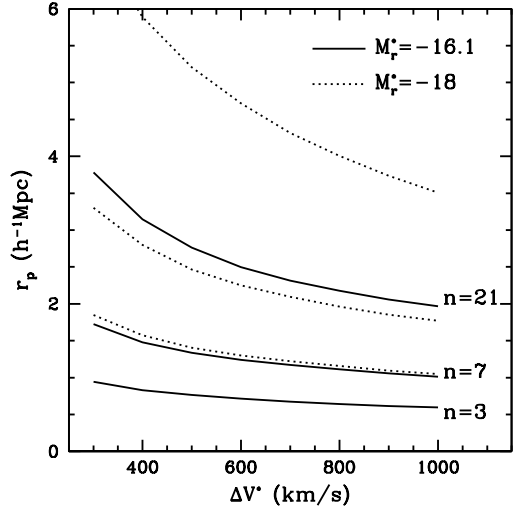}
%\vspace{-1.2cm}
\caption{Projected distance to the $n$th nearest neighbor $r_{p}$ as a
function of $\Delta V^{*}$. The line types distinguish the
luminosity constraint: $M^{*}=-16.1$ (solid line) and $M^{*}=-18$ (dotted line).
The number of nearest neighbors $n$ is indicated for the case of $M^{*}=-16.1$.
}
\end{figure}
%%%%%%%%%%%%%%%%%%%%%%%%%%%%%%%%%%%%%%%%%%%%%%%%%%%%%%%%%%%%%%%%%%%%%%%%%%%%

Since $r_{p}$ characterizes the length scales relevant to the
local background densities,
the environmental properties represented by the local
background densities derived from the 3rd nearest neighbor with $M^{*}=-18$
are equally well described by those from the 7th nearest neighbor with
$M^{*}=-16.1$. The same is true for the mean $r_{p}$ derived by $n=21$ with
$M^{*}=-16.1$ and $n=7$ with $M^{*}=-18$.
Because we used local background densities normalized by their
mean values, the frequency distributions of the local background densities
derived from the two methods are expected to be nearly identical.
In the literature (Muldrew et al. 2012, references are there in),
several values of $n$ were used to derive local background
densities. However, as seen in Fig. 4, a large $n$ such as $n=21$ should
be used for the structures with length scales larger than $\sim4$ $~h^{-1}$Mpc,
while a small $n$ such as $n=3$ appears to be suitable for systems smaller
than $\sim1$ $~h^{-1}$Mpc. For small systems such as galactic satellite
systems or group of galaxies,
$n\leq5$ with $\Delta V^{*}\approx$ 500~km/s appears to be a
better choice for the local background densities.

\subsection{Local Background Density from the $n$th Nearest Neighbor}

Currently, the $n$th nearest neighbor technique is most widely used method
to derive the local background density. It has been applied to nearby
galaxies as well as to large scale structures.
The local background density obtained using the $n$th nearest neighbor
is defined as

\begin{equation}
\Sigma_{n}= {n \over {4\pi {r_{p,n}}^{2}}},
\end{equation}

\noindent{where $r_{p,n}$ is the projected distance to the $n$th nearest
galaxy that is brighter than $M^{*}$ with $|\Delta V|<\Delta V^{*}$.}
We use $\Delta V^{*}$=~500~km/s and $M^{*}=-16.1$ for the representative
peculiar velocity and luminosity constraints.
%%%%%%%%%%%%% fig5: fig5Cdv5m16.eps %%%%%%%%%%%%%%%%%%%%%%%%%%%%%%%%%%
\begin{figure}[!t]
\center
\includegraphics[scale=0.6]{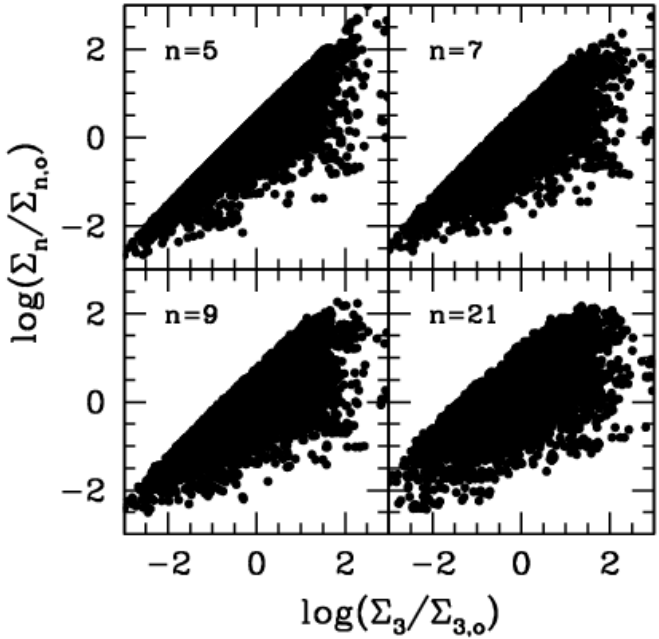}
\caption{
Correlation between background densities derived from
the $n$th nearest neighbor method. We selected densities from $n=3$ as the
reference densities. We fixed the two
searching parameters: $\Delta V^{*}$=~500~km/s and $M^{*}=~-16.1$.}
\end{figure}
%%%%%%%%%%%%%%%%%%%%%%%%%%%%%%%%%%%%%%%%%%%%%%%%%%%%%%%%%%%%%%%%%%%%
However, other values of $\Delta V^{*}$ and $M^{*}$ are used
together to examine the dependence of the density distributions
on $\Delta V^{*}$ and $M^{*}$. In the literature,
as discussed by \citet{mul12}, there is no consensus on the
number of neighbor galaxies to be searched. It ranges from 3 to 60 with
a preference for small $n$ values for local background density
relevant to the fine morphologies of galaxies. A large $n$ such as $n=21$
appears to be a better measure of the local background density relevant to
large scale structure \citep[see][for example]{par08}. Fig.~5 shows the
correlations between the densities derived from $n=5, 7, 9, 21$ and that from
$n=3$. We maintain the searching constraints $\Delta V^{*}$ and $M^{*}$
constant at 500~km/s and -16.1, respectively. There are fairly good correlations
between the densities derived from different $n$, although the scatter is
too large to be ignored. The scatter is mainly due to the presence of
small scale regions with different densities $\Sigma_{3}$ within regions of
a density $\Sigma_{n}$ for a large $n$.
As expected from the dependence of scale lengths on $n$, the
densities from
$n=5$ exhibit the best correlation with $n=3$, while those from
$n=21$ exhibit the worst correlation.

Fig.~6 shows the fractional distribution of densities derived from the $n$th
nearest neighbor with $M^{*}$=-16.1 and -18 as well as $\Delta V^{*}=$~500 and
1000~km/s. There is no significant difference between the
fractional distributions for the same $n$. The density
distribution pattern is less affected by $M^{*}$ than by $\Delta V^{*}$.
However, the case for $n=3$ with $\Delta V^{*}=$~500~km/s
and $M^{*}=-16.1$ exhibits the largest density ranges, which results in the
highest resolution. For the same $\Delta V^{*}$ and $M^{*}$,
the density ranges decrease from the smallest to the largest $n$ with a
peak density shift towards higher densities. The shift in the peak density with
larger $n$ is due to the larger contribution of dense regions such as
clusters of galaxies for larger $n$.
The dependence of density resolution on $n$
is due to the characteristic scales associated with
the derived densities. As demonstrated in Fig.~4, the mean
$r_{p}$ depends on $n$ if we fix $M^{*}$ and $\Delta V^{*}$.
When compared with the densities from the fixed aperture method,
the ranges of the
densities derived from the $n$th nearest neighbor are larger than those
from the fixed apertures.
This suggests that if morphological features
such as spiral arm classes are mostly affected by small scale
perturbations, the most relevant density estimator would be the one derived
from the 3rd nearest neighbor with $\Delta V^{*}$=~500~km/s.

%%%%%%%%%%%%% fig6: fig6rhon4.eps %%%%%%%%%%%%%%%%%%%%%%%%%%%%%%%%
\begin{figure}[!t]
\center
\includegraphics[scale=0.45]{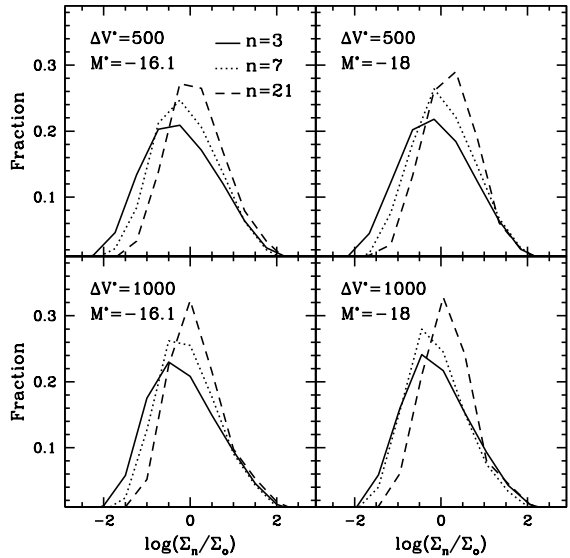}
%\vspace{-1.2cm}
\caption{Fractional distribution of background densities obtained using
the $n$th nearest neighbor.
Left and right panels present
the densities derived using $M^{*}=-16.1$ and $M^{*}=-18$, respectively.}

\end{figure}
%%%%%%%%%%%%%%%%%%%%%%%%%%%%%%%%%%%%%%%%%%%%%%%%%%%%%%%%%%%%%%%%%%%%%%%

\subsection{Disk Morphology and Local Background Densities}

The arm classes, Hubble type, and bar type fractions of 1912
spiral galaxies as a function of local background densities are shown in
Figs.~7, 8 and 9. The local background densities in Figs.~7 and 9 were
derived using $n=3$, while in Fig.~8 they were derived using $n=5$.
We use $\Delta V^{*}$=500 km/s in all the cases except in the lower panels
of Fig.~7 where $\Delta V^{*}$=~1000~km/s.
We used the same luminosity constraint of $M^{*}$=~-16.1 in Figs.~7 and
8; however, we used $M^{*}$=~-18 in Fig. 9 to observe the effects of the
luminosity constraint. We used the arm class, Hubble type, and bar type of 1912 spiral galaxies
from the visual classification of \citet{ann13}.
There is a strong dependence of arm class and Hubble type on the local
background densities; however,
the bar type exhibits a very weak dependence on the local background densities.

%%%%%%%%%%%%%%%  Fig7: atbm16n3.ps (rhoatb.ps) %%%%%%%%%%%%%%%%%%
\begin{figure}[!t]
\center
\includegraphics[scale=0.4]{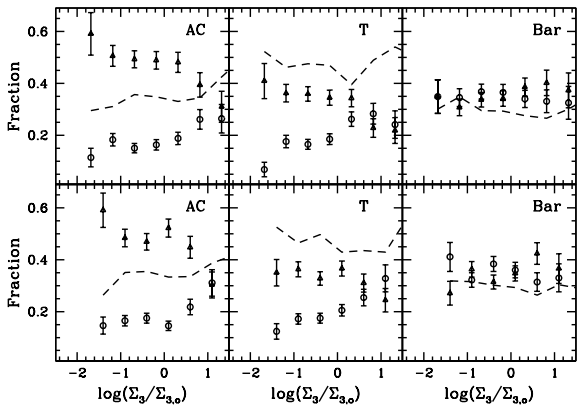}
 \vspace{-0.02cm}
\caption{Fraction of disk morphology as a function of the local
background density ($\Sigma_{3}$) derived from $\Delta V^{*}$=~500~km/s (upper
panels) and $\Delta V^{*}$=~1000~km/s (lower panels). We fixed $M^{*}$ to
$M_{r}=~-16.1$. The fractions of the arm class, Hubble type, and bar type are plotted
in the left, middle, and right panels, respectively.
The open circles with error bars
represent the grand design in the left panels, early-type spirals in the middle
panels, and SA in the right panels. The
filled triangles with error bars represent the
flocculent arms in the left panels, late-type spirals in the middle panels, and
SB spirals in right panels. The short dashed lines represent multiple-arm
in the left panels, intermediate-types in the middle panels, and SAB in
the right panels.}
\end{figure}
%%%%%%%%%%%%%%%%%%%%%%%%%%%%%%%%%%%%%%%%%%%%%%%%%%%%%%

As shown in the left panels of Figs. 7, 8, and 9, the fraction of grand design
galaxies (open circles with error bars) steadily increases from low density
regions to high density regions, while that of flocculent galaxies (filled
triangles with error bars) decreases.
Since structural properties are very
different, the opposite trends in grand design spirals and flocculent spirals
demonstrates the dependence of arm class on the local background density.
The fraction of multiple-arm galaxies (dashed lines),
which have intermediate arm properties between those of grand design and
flocculent arms, is nearly constant along the
local background density. The correlations between arm class fraction and
local background densities is not affected much by the neighbor search
constraint $\Delta V^{*}$ for $n=3$ and $n=5$ when we use the luminosity
constraint $M^{*}$ as the limiting magnitude of the volume limited
sample. However, for the case of $M^{*}$=-18, the correlation
becomes too weak to be easily recognized. The clearest correlation is
observed for the local background densities derived from $n=3$ with
$\Delta V^{*}$=~500~km/s and $M^{*}$=~-16.1.

For the Hubble types plotted in the middle panels of Figs.~7, 8, and 9, the
fractions of early-type spirals (open circles with error bars) increase
with increasing local background density, while those of late-type
spirals (filled triangles with error bars) decrease with increasing local
background density. The correlation of the early-type fractions with local
background density is stronger than that of the grand design fractions. However,
the anti-correlation of the late fractions with
local background density is weaker than that of the flocculent arms,
especially for the densities with $\Delta V^{*}$=~1000~km/s, regardless of
$n$ and $M^{*}$. Because intermediate-type spirals (dashed lines) have intermediate
morphological properties
between those of early- and late-type spirals, we expect that the
correlation between their fractions and local background densities are
also intermediate; in other words, we expect almost no correlation. In this
regard, the nearly constant fractions of intermediate-type spirals,
shown in the upper middle panel in Fig.~7 and middle panel in Fig.~8,
appear to be consistent with the general MDR.
The decreasing trend of intermediate spiral fractions
shown in the lower panels of Fig.~7 appears plausible because
the late-type fractions also decrease with density at a similar rate. For the
densities from $n=3$ and $M^{*}$=~-18 shown in Fig.~9, the correlations between the
disk morphology and $\Sigma_{3}$ with $M^{*}$=~-18 are much weaker than
those shown in Figs.~7 and 8.

%%%%%%%%%%%%%%%  Fig8: fig8atbm16n5.eps %%%%%%%%%%%%%%%%%%%%%%
\begin{figure}[!t]
\center
\includegraphics[scale=0.4]{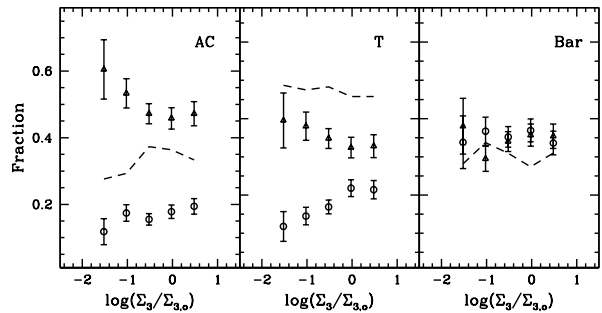}
\caption{Fraction of disk morphology as a function of local
background density ($\Sigma_{5}$) derived from $\Delta V^{*}$=~500~km/s.
We fixed $M^{*}$ as $M_{r}=-16.1$. The meanings of symbols
and line types are the same as those of Fig. 7.
}
\end{figure}
%%%%%%%%%%%%%%%%%%%%%%%%%%%%%%%%%%%%%%%%%%%%%%%%%%%%%%

\citet{ann13} determined the dependence of the disk morphology on the local
background density, calculated using the nearest neighbor method
with $n=5$. Here, we present the same dependence with $n=3$ (Fig.~7) because the
local background density derived from $n=3$ exhibits the highest resolution
owing to its large density range (Fig.~6). As shown in the left panel of Fig.~7,
the fractions of arm classes are highly dependent on the local
background density. The fraction of flocculent galaxies (filled triangles)
is $\sim0.6$ at ${\rm log} (\Sigma_{3}/\Sigma_{3,0}) = -2$ and decreases
monotonically as the local background density increases, reaching $\sim0.3$ at
${\rm log} (\Sigma_{3}/\Sigma_{3,0}) = 1$. Conversely,
the fraction of grand design galaxies (open circles) increases nearly
monotonically from $\sim0.1$ at ${\rm log} (\Sigma_{3}/\Sigma_{3,0}) =-2$ to
$\sim0.35$ at ${\rm log} (\Sigma_{3}/\Sigma_{3,0}) =1$. Multiple-arm
galaxies (dashed line) exhibit an
increasing trend similar to that of grand design galaxies, but with a much smaller
gradient. The general trend of the dependence of arm class on the local
background density shown in Fig.~7 agrees well with that given in \citet{ann13},
but there is a slight difference between them because of the different density
resolution resulting from the different $n$ used.
The dependence of the Hubble type on the local background density is
stronger than that of the arm class. This is because the bulge-to-disk ratio,
which is strongly correlated with the Hubble type, is determined during the
galaxy formation process with a large bulge-to-disk ratio in high density
regions following the MDR. Conversely, the arm class is mainly determined by the
mode of the disk instability, which depends on the external tidal perturbation
as well as the disk mass relative to halo mass.

%%%%%%%%%%%%%%  Fig9: fig9atbm18n3.eps %%%%%%%%%%%%%%%%%%%%%%
\begin{figure}[!t]
\center
\includegraphics[scale=0.4]{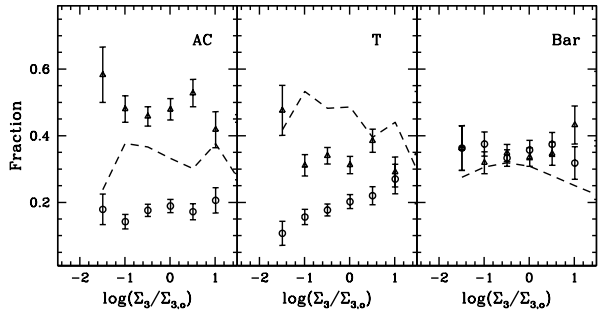}
\caption{Fraction of disk morphology as a function of the local
background density ($\Sigma_{3}$) derived from $\Delta V^{*}$=~500~km/s.
We fixed $M^{*}$ as $M_{r}=~-18$.
The fractions of arm class, Hubble type, and bar type are plotted
in the left, middle, and right panels, respectively. The meanings of symbols
and line types are the same as those of Fig. 7.
}
\end{figure}
%%%%%%%%%%%%%%%%%%%%%%%%%%%%%%%%%%%%%%%%%%%%%%%%%%%%%%

\section{OTHER MEASURES OF LOCAL ENVIRONMENT}

\subsection{Projected Distance to the Nearest Neighbor}

The projected distance ($r_{p}$) appears to be a good measure for the galaxy
environment because the morphology of a galaxy is mostly affected by its
neighbor
galaxies \citep{par08}. The strong dependence of morphology conformity
between a host galaxy and its satellite galaxies \citep{wei06,ann08} is a
good example of the role $r_{p}$ has on the galaxy morphology.
Sometimes, $r_{p}$
is normalized by the virial radius of the neighbor galaxy because the morphology
transformation driven by interactions with neighbor galaxies is effective
when galaxies are located within the virial radii of their neighbor galaxy
\citep{par09}.

Fig. 10 shows the arm class, Hubble type, and bar type fractions as functions
of $r_{p}$; these were derived with $\Delta V^{*}$=~500~km/s and $M^{*}$=~-18.
It is apparent that the arm class, Hubble
type, and bar type do depend on $r_{p}$. The fraction of grand design spirals
decreases with increasing $r_{p}$,
while flocculent spirals exhibit the opposite
trend. The fraction of multiple-arm spirals is nearly constant until
$r_{p}$=~1~$~h^{-1}$Mpc and decreases thereafter. For the Hubble type case,
the fraction of early-type spirals decreases with $r_{p}$ and that of late-type
spirals increases with $r_{p}$. Meanwhile, the bar type also
exhibits a clear dependence on $r_{p}$,
in the sense that the SA fraction increases with increasing $r_{p}$ and
the SB fraction decreases with $r_{p}$,
while the SAB exhibits a nearly constant fraction
along $r_{p}$. Thus, $r_{p}$ appears to be a better measure of the local
environment to distinguish the disk morphology,
especially arm classes, than the local background density.

%%%%%%%%%%%%%  Fig10: fig10atbrpMpcv5m18.eps %%%%%%%%%%%%%%%%%%
\begin{figure}[!t]
\center
\includegraphics[scale=0.4]{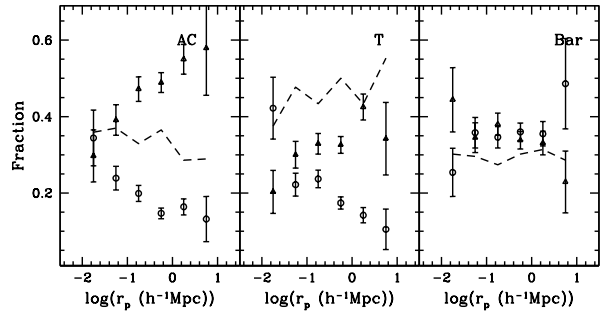}
\caption{Arm class, Hubble type, and bar type fractions as functions of
$r_{p}$ with $M_{r}=-18$. The meanings of symbols
and line types are the same as those of Fig. 7.
}
\end{figure}
%%%%%%%%%%%%%%%%%%%%%%%%%%%%%%%%%%%%%%%%%%%%%%%%%%%%%%

Correlations between disk morphology
and $r_{p}$ are stronger than those between disk morphology and local
background density (Figs. 7 and 8). In particular, $r_{p}$ exhibits much
stronger correlations with grand design spirals and early-type spirals
than for the local background densities (Figs. 7 and 8). However, if we
use $r_{p}$ derived with $M_{r}$=~-16.1, correlations become much weaker,
as shown in Fig. 11. The reason for stronger correlations for $r_{p}$ with
$M_{r}$=~-18 is the dependence of the interaction strength on the neighbor
luminosity, i.e., mass. This explanation is supported by the strongest
dependence of disk morphology on the tidal index ($TI$) shown in Fig.~13.

Fig. 12 shows the dependence of the disk morphology on the projected distance
normalized by the nearest neighbor's virial radius ($r_{virnei}$),
$r_{p}/r_{virnei}$. The general dependence is similar to that shown
in Fig.~10 However, the arm class exhibits a better correlation
with $r_{p}$ than $r_{p}/r_{virnei}$, while the Hubble type exhibits a
better correlation with $r_{p}/r_{virnei}$ because there is a break in the
grand design fraction at small normalized projected separations of  
$r_{p}/r_{virnei} \approx0.1$. Meanwhile, a similar break occurs in the early-type fractions
at small projected separations of $r_{p} \approx 0.1 h^{-1}Mpc$. The grand design fraction is nearly constant at $r_{p}  > 1h^{-1}Mpc$ 
or $r_{p}/r_{virnei} > 1$. The
presence of a break in the grand design fraction at
$r_{p}/r_{virnei} \approx 0.1$ and the absence of a break in the early-type
fraction at the same $r_{p}/r_{virnei}$ suggest that the disk instability is
affected more by the close neighbors than the global structure of disk, which is
represented by the bulge-to-disk ratio. The nearly constant fraction of the grand design
spirals at $r_{p}/r_{virnei} \approx 10$ implies that the length scales
on which the arm class depends are smaller than those of the Hubble type
because early-type fractions exhibit a nearly monotonic decrease to
$r_{p}/r_{virnei} \approx 10$.
There is little difference in the dependence of bar fractions between $r_{p}$
and $r_{p}/r_{virnei}$.

%%%%%%%%%%%%%%%%%%%%%%%%%%%%%%%%%%%%%%%%%%%%%%%%%%%%%%%%%%%%%%%%%%%%%%
\begin{figure}[!t]
\center
\includegraphics[scale=0.4]{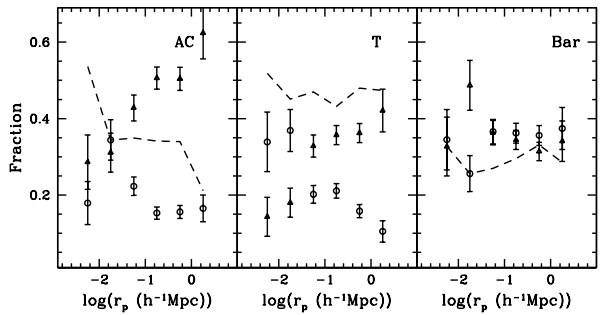}
\caption{The arm class, Hubble type, and bar type fractions as functions of
$r_{p}$ with $M_{r}=~-16.1$. The arm class, Hubble type, and bar type
are plotted in the left, middle, and right panels, respectively.
The meanings of symbols
and line types are the same as those of Fig. 7.
}
\end{figure}
%%%%%%%%%%%%%%%%%%%%%%%%%%%%%%%%%%%%%%%%%%%%%%%%%%%%%%

\subsection{Tidal Index}

Tidal interactions with neighbor galaxies are considered to be the main
drivers of morphology change of a galaxy.
A strong correlation of disk morphology with the local background density
suggests a parameter that represents the tidal strength of neighboring
galaxies can be used as a measure of the local environment of galaxies.

%%%%%%%%%%%%%  Fig12: fig12atbs2rv5m18.eps %%%%%%%%%%%%%%%%%%
\begin{figure}[!]
\center
\includegraphics[scale=0.4]{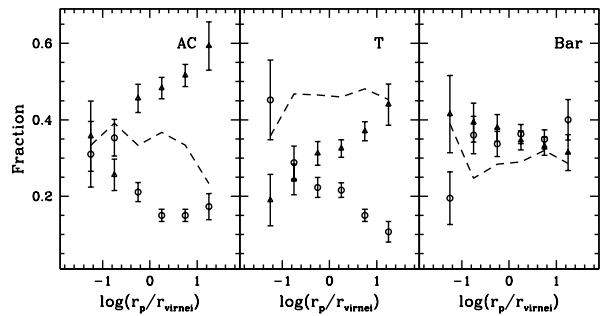}
\caption{The arm class, Hubble type, and bar type fractions as functions of
$r_{p}/r_{virnei}$. The arm class, Hubble type, and bar types are plotted
in the left, middle, and right, respectively. The meanings of symbols
and line types are the same as those of Fig. 7.
}
\end{figure}
%%%%%%%%%%%%%%%%%%%%%%%%%%%%%%%%%%%%%%%%%%%%%%%%%%%%%%

\citet{kar99} introduced the $TI$ as a measure of the tidal interaction
among galaxies in the local volume ($D < 10$ Mpc). They used the three-dimensional
separation between neighboring galaxies based on the distances, most of which are
determined by direct methods such as photometric distances or group membership
with known group distances. However, we cannot use the three-dimensional
separation of neighboring galaxies owing to the large uncertainties that are inherent
in the kinematic distances derived from the galaxy redshifts.
We defined a $TI$ similar to \citet{kar99} using the projected
separation between a target galaxy $i$ and its neighbor galaxies as

\begin{equation}
TI_{i}= max[log ({Gm_{k} \over {r_{ik}^{3}}}) + C],  k=1, 2, 3, ....N
\end{equation}
\noindent{where $m_{k}$ and $r_{ik}$ are, respectively, the mass and projected separation
between the target galaxy $i$ and neighbor galaxy $k$. $C$ is the normalization
constant, which corresponds to the $TI$ using a galaxy containing the Milky Way mass at
a separation of 1~$~h^{-1}$Mpc.}
We used galaxy masses determined from $M_{r}$ using a mass-to-light
ratio of 1.

Fig. 13 shows the dependence of disk morphology on the $TI$. Here, we used the
neighbor search constraints $\Delta V^{*}$=~500~km/s and $M^{*}$=~-16.1. As
expected, the arm class and Hubble type clearly depend on $TI$.
The fractions of grand design and multiple-arm spirals increase with $TI$,
while that of the flocculent spirals decreases with $TI$. This means that
strong tidal interactions are likely to drive a large scale disk instability,
which leads to generation of global density waves. However, this general trend
is kept at moderate ranges of $TI$. The flat distribution of the
grand design fractions at $TI < 0$ suggests that the tidal strength weaker than
that caused by a galaxy with a Milky way mass at a separation
of $\sim1$ $~h^{-1}$Mpc does not generate global density waves to drive grand
design arms. The grand design fraction also decreases at extremely strong
tidal forces, $TI > 5$. This is due to the formation of a single massive arm or
chaotic arms, which are classified as flocculent spirals.
For the Hubble type case, the fraction of early-type spirals increases with
$TI$, while that of late type spirals decreases with $TI$ and exhibit a bump at
$TI \approx 3$. Meanwhile, the intermediate-type spirals exhibit nearly constant fractions.

Contrary to the cases for the local background density and projected
distance to the nearest neighbor, the velocity constraint of neighbor galaxies
is much tighter.
As shown in Fig. 14 in which we derived $TI$ with
$\Delta V^{*}$=~1000~km/s and $M^{*}$=-16.1, the dependence of disk morphology
on $TI$ is much weaker than that in Fig. 13. In particular, there appears
to be no correlation between the arm class and $TI$ for $TI < 4$. A similar trend
is observed for the Hubble type with $TI < 4$. This suggests that
$\Delta V^{*}$=~1000~km/s allows too many interlopers in the neighbor galaxies.

%%%%%%%%%%%%%  Fig13: fig13atbTIv5.eps %%%%%%%%%%%%%%%%%%%%%%%
\begin{figure}[!t]
\center
\includegraphics[scale=0.4]{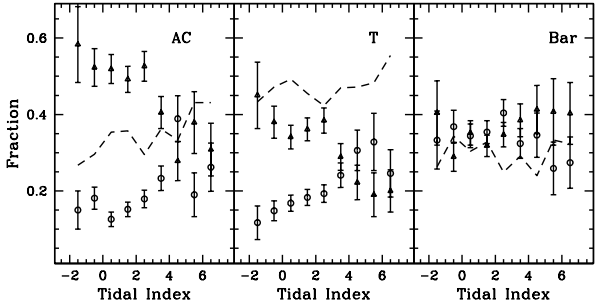}
\caption{The arm class, Hubble type, and bar type fractions as functions of
$TI$. Here, $TI$ is derived with $\Delta V^{*}$=~500~km/s.
The arm class, Hubble type, and bar types are plotted
in the left, middle, and right panels, respectively. The meanings of symbols
and line types are the same as those of Fig. 8.
}
\end{figure}
%%%%%%%%%%%%%%%%%%%%%%%%%%%%%%%%%%%%%%%%%%%%%%%%%%%%%%

\section{DISCUSSION}

We explored three measures of local environment: local background
density ($\Sigma_{n}$), projected distance to the nearest neighbor ($r_{p}$),
and $TI$ which quantifies the maximum tidal strength from neighbor galaxies.
These measures exhibit fairly good correlations with disk morphology
characterized by arm class (G, M, F), Hubble type (early, intermediate, late)
and bar type (SA, SAB, SB).
However, the environment measures are heavily affected by the neighbor search
parameters ($M^{*}$ and $\Delta V^{*}$) that constrain
the neighbor's luminosity and velocity relative to target galaxy.

It is natural to take $M^{*}$ as the limiting luminosity
that defines the present sample of galaxies as a volume limited sample,
i.e., $M^{*}$=~-16.1. However, we examined other values of $M^{*}$.
As expected, we found the best correlations between the disk morphology and
environment measures ($\Sigma$ and $TI$) when we used $M^{*}$=~-16.1; however,
$M^{*}$=~-18 is better for $r_{p}$. In particular, the arm class exhibited
the best correlation with $r_{p}$ when it is derived with $M^{*}$=~-18.
There is no preference for $\Delta V^{*}$ because the particular velocities,
which determine $\Delta V^{*}$, depend on the environment itself.
Too small values of $\Delta V^{*}$ cause a large fraction of neighbors to
be missed, while too large values of $\Delta V^{*}$ introduce a large number
of interlopers that degrade the correlation between the disk morphology
and environment.
However, as shown in Fig. 3, the dependence of $r_{p}$ on $\Delta V^{*}$ is
very weak, especially for $M^{*}$=~-16.1.
%%%%%%%%%%%%%  Fig14: fig14atbTIv10.eps %%%%%%%%%%%%%%%
\begin{figure}[!t]
\center
\includegraphics[scale=0.4]{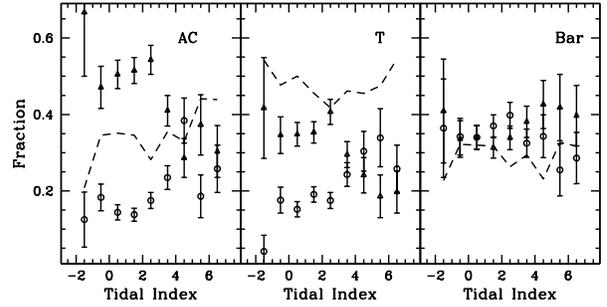}
\caption{The arm class, Hubble type, and bar type fractions as functions of
$TI$. Here, $TI$ is derived using $\Delta V^{*}$=~1000~km/s.
The arm class, Hubble type, and bar types are plotted
in the left, middle, and right panels, respectively. The meanings of symbols
and line types are the same as those of Fig. 8.
}
\end{figure}
%%%%%%%%%%%%%%%%%%%%%%%%%%%%%%%%%%%%%%%%%%%%%%%%%%%%%%
We extensively examined two cases of $\Delta V^{*}$: 500 and 1000~km/s.
The 500~km/s value of $\Delta V^{*}$ is widely used to search for nearby
neighbors such as satellite galaxies (for example see \citet{ann08}).
It is similar to those used by \citet{par09} to search for the nearest
neighbor. $\Delta V^{*}$ values used by \citet{par09} were 400~km/s for
neighbors of spiral galaxies and 600~km/s for neighbors of elliptical galaxies.
The 600~km/s value of $\Delta V^{*}$ is a typical value for selecting
isolated galaxies; it was used in, for instance, \citet{par09}.
We found that better correlations between the disk morphology and measures of
local environment are observed when we use $\Delta V^{*}$=~500~km/s.
However, as demonstrated in Fig. 7, correlations
between the disk morphology and $\Sigma_{n}$ are not affected much by
$\Delta V^{*}$ for small $n$ values such as $n=3$ or $n=5$. The preference
for a $\Delta V^{*}$ value of 500~km/s is more pronounced in the dependence
of disk morphology on $TI$, which is thought to be a direct
measure of tidal strength of neighboring galaxies. Tighter constraints on
$|\Delta V|$ results in a better correlation between the disk morphology
and environment measures because interlopers are suppressed.

Arm classes exhibit stronger dependencies on $r_{p}$ than $\Sigma$ and $TI$,
while Hubble types exhibit stronger dependencies on $\Sigma$. The strong
dependence of the arm class on $r_{p}$ is due to the close relationship
between the arm class and mode of disk instability, which is thought to be
determined primarily by the perturbations imposed by the nearest neighbor.
The less clear correlation between the arm class and $\Sigma$ could be
due to multiple criteria of the arm class (continuity, symmetry, and length)
that degrade correlations. Some arms are classified as a flocculent galaxy
owing to the lack of symmetry but have continuity, while other arms are
classified as a flocculent galaxy owing to the fleece-like structure.
A galaxy with two symmetric
arms can be classified as a grand design galaxy if there is no other structure,
but it can be classified as a multiple-arm galaxy if there is an outer ring or a
pseudo-ring. Thus, the arm class of multiple arms can be merged
with grand design arms if we consider only symmetry and continuity. On the
contrary, the Hubble type subdivision of spiral galaxies, i.e., Hubble
stage, is basically classified by a single parameter, bulge-to-disk ratio, which
obeys the MDR.

It is also possible that the correlation between the Hubble type and local
background density is more fundamental than that between the arm class and
local background density. The Hubble type, which is distinguished by
the bulge-to-disk ratio, is determined during the formation processes of
the bulge and disk, while the arm class, which is distinguished by
the spiral arm structures and determined by disk instability, is set up
after the disk forms. As dictated by MDR \citep{dre80}, which is closely
related with the morphology-luminosity relation (MLR) of
galaxies \citep{par07}, bulge dominated galaxies are
likely to be formed in dense regions. An elliptical galaxy is an
extreme case of a bulge dominated galaxy. The bulge of the early-type spiral is
also thought to be formed in dense regions where a rapid collapse of the
central region is likely to occur before the disk forms.
However, in less dense regions, the central regions slowly collapse
owing to the dependence of the dynamical time scale on the density,
$\tau \sim (1/G\rho)^{1/2}$.
Thus, it appears that it is better to use the bulge-to-disk ratio
as the primary parameter to distinguish the Hubble stage. It can be supported
by the fact that the openness of the spiral arm depends on the bulge mass.
Large bulges drive tightly wound spiral arms, while small bulges drive
loosely wound spiral arms.

It seems of interest to clarify whether the bar type depends on the local
environment. There have been many investigations regarding the relation
between the bar fraction and environment (see \citet{lee12} and
references therein).
The nature versus nurture controversy for bars in disk galaxies is
one of the long-lasting issues in galaxy evolution.
The present analysis does not yield a clear conclusion, but we do obtain
some insights. As shown in Figs. 7, 8, and 9, there is no strong
bar type dependence on the local background density, especially in the
galaxies located in the moderate density regions. However, there is a weak trend
of increase in the fraction of SB galaxies with increasing density and decrease in the fraction of
SA galaxies with increasing density (upper right panel of Fig. 7)
for $\Sigma_{3}$,
especially those derived from $\Delta V^{*}$=~500~km/s. Since $\Sigma_{3}$
with $\Delta V^{*}$=~500~km/s and $M^{*}$=-16.1 provide a high density
resolution and strong correlations with the arm class and Hubble type,
it seems plausible that the unsolved issue of the nature versus nurture
controversy is partly due to a lack of density resolution required
for a detailed
analysis. Of course, it is also plausible that the recurrent formation and
destruction of bars demolish the correlation between the bar
fraction and environment. However, if bars are robust enough to survive
the internal and external destruction mechanisms,
the controversy may be settled if we devise a
better measure of galaxy environment. This argument is supported by better
correlations between the bar fraction and $TI$, which is the most direct
measure of the tidal strength from neighboring galaxies.

\section{CONCLUSIONS}

It is apparent that the disk morphology is closely related with the local
environment of galaxies. Arm class and Hubble type exhibit a clear dependence
on three measures of the local environment: local background density ($\Sigma$),
projected distance to the nearest neighbor galaxy ($r_{p}$), and tidal
index ($TI$). However, the correlation strength depends on the environmental
measures used. Since measures of the galaxy environment are constrained
by the neighbor search parameters, $M^{*}$ and $\Delta V^{*}$, correlation
strengths also depend on $M^{*}$ and $\Delta V^{*}$. It was found that
$\Delta V^{*}$=~500~km/s is better than $\Delta V^{*}$=~1000~km/s for all
three measures, $\Sigma$, $r_{p}$, and $TI$, but $M^{*}$ can be different
for the measures of the selected environment.

Arm class exhibited the strongest correlation with $r_{p}$ when $r_{p}$ is
derived
from the neighbor search constraints $M^{*}=$~-18 and $\Delta V^{*}$=~500~km/s.
The fraction of grand design spirals decreases with increasing $r_{p}$, while
that of flocculent spirals increases with it. Multiple-arm spirals exhibit a
nearly constant fraction for $r_{p} <$ 1 $~h^{-1}$Mpc and decreases as $r_{p}$ increases further. $TI$ as well as $\Sigma_{n}$ with $n=3$ and $5$ are also
closely related with the arm classes of spiral galaxies; however,
they exhibit a weaker correlation than $r_{p}$. It is worth noting
that the luminosity constraint of $r_{p}$ that gives strongest
correlation with arm class is $\sim2$ magnitude brighter than the limiting
magnitude that defines the present volume limited sample,  $M^{*}$=~-16.1.
This means that the arm class is mostly affected by a massive neighbor galaxy,
which is close enough to significantly perturb the disk instability of
target galaxy.

The local background density, $\Sigma_{n}$, is most closely related with
the Hubble
stage of spiral galaxies; however, $r_{p}$ and $TI$ also exhibit fairly good
correlations with Hubble type. The best correlation was found for
$\Sigma_{5}$ with $M^{*}=-16.1$ and $\Delta V^{*}$=~500~km/s.
The fraction of early-type spirals increases with increasing $\Sigma$
and $TI$ as well as decreasing $r_{p}$,
while that of late-type spirals exhibited opposite trends.
Intermediate-type spirals exhibited a nearly constant fraction along $\Sigma$,
$TI$, and $r_{p}$. The luminosity constraint of $r_{p}$ was $M^{*}$=~-18, while
that for $\Sigma$ and $TI$ was $M^{*}$=-16.

The bar type exhibited a weak dependence on the three environment measures, but there was
a high frequency of strongly barred galaxies at high density
regions where we expect relatively strong tidal strength and small
separations of neighbor galaxies. The opposite is true for non-barred
galaxies. Thus, the present results support the idea that bar formation is
affected by the environment of galaxies.

The strong correlations between the frequencies of Hubble type of
spiral galaxies, i.e., Hubble stage,
and the local background density appear to originate
from the MDR and MLR of galaxies, which govern the galaxy morphology throughout
the galaxy formation process. This is the reason why the Hubble type exhibits
a fairly strong correlation with $\Sigma$. The strong dependence of
the frequencies of the arm class on the projected distance to the nearest
neighbor with the neighbor luminosity brighter than $M^{*}$=~-18 suggests that
spiral arm structures, which are distinguished by arm class, are most
affected by local perturbations caused by the nearest
massive neighbor galaxy. If the perturbation is
strong enough to drive the global density wave in the disk, grand design spiral
arms are formed. However, if the perturbations are too strong, they break the symmetry of spiral
arms and produce flocculent arms such as a single massive spiral arm.

\acknowledgments

{This work was supported for two years by Pusan
National University Research Grant.}

\end{document}